\documentclass[prl,aps,groupedaddress,10pt,floatfix,twocolumn,amsmath,amssymb]{revtex4-1}

\usepackage{graphicx}
\usepackage{dcolumn}
\usepackage{bm}
\usepackage{newfloat}
\DeclareFloatingEnvironment[name={Figure S}]{suppfigure}

\makeatletter 
\def\tagform@#1{\maketag@@@{(S\ignorespaces#1\unskip\@@italiccorr)}}
\makeatother
 
\makeatletter
\makeatletter \renewcommand{\fnum@suppfigure}
{\figurename~S\thesuppfigure}
\makeatother
 
\renewcommand{\figurename}{Figure}

\begin{document} 

\title{Observation of Radiation Pressure Shot Noise on a Macroscopic Object} 
\author{T. P. Purdy}
\email{tpp@jila.colorado.edu}
\author{R. W. Peterson}
\author{C. A. Regal}
\affiliation{JILA, University of Colorado and National Institute of Standards and Technology,}
\affiliation{and Department of Physics, University of Colorado, Boulder, Colorado 80309, USA}

\date{\today}

\begin{abstract}
\bf{The quantum mechanics of position measurement of a macroscopic object is typically inaccessible because of strong coupling to the environment and classical noise.  Here we monitor a mechanical resonator subject to an increasingly strong continuous position measurement and observe a quantum mechanical backaction force that rises in accordance with the Heisenberg uncertainty limit.  For our optically-based position measurements, the backaction takes the form of a fluctuating radiation pressure from the Poisson-distributed photons in the coherent measurement field, termed radiation pressure shot noise.  We demonstrate a backaction force that is comparable in magnitude to the thermal forces in our system.  Additionally, we observe a temporal correlation between fluctuations in the radiation force and in the position of the resonator.}   
\end{abstract}

\maketitle 

	In measuring the trajectory of an object at the scale of our everyday experience we rarely consider the fundamental limitations imposed by quantum mechanics. Yet quantum-mechanical effects are present even when monitoring the position of macroscopic objects and are expected to soon limit, for example, the precision of gravitational wave observatories~\cite{Harry10}. Imagine measuring the position of an object to an accuracy $\Delta x$. A momentum uncertainty of at least $\Delta p = \hbar/2\Delta x$ must then be present, where $\hbar$ is the reduced Planck's constant that appears in the Heisenberg uncertainty relation. This requisite momentum (or equivalently velocity) uncertainty adds position uncertainty at a later time. Thus, an observer must weigh pinpointing the location of the object against introducing quantum measurement backaction that obscures the subsequent motion. 
	
	For an optical position measurement, this quantum backaction is termed radiation pressure shot noise (RPSN) \cite{Braginsky78,Caves81}.  Here a fluctuating force arises from, for example, the recoil momentum transfer of randomly arriving photons (shot noise) reflecting off an object.  In the next generation advanced gravitational wave observatories, such as LIGO \cite{Harry10}, Virgo, and KAGRA \cite{Somyia12}, RPSN is predicted to limit sensitivity even with tens of kilograms test masses.  Ideas developed to circumvent quantum limits imposed by backaction include quadrature-squeezed light \cite{Kimble01} and backaction evasion techniques \cite{Braginsky80,Somyia12}.  However, for typical objects, the scale of quantum backaction is small compared to thermal motion or classical probing noise.  In this report, we observe RPSN on a solid macroscopic (visible to the naked eye) mechanical resonator by using an optical interferometric measurement of its vibrational motion.


	Figure 1A shows the canonical picture of a Heisenberg-limited continuous position measurement.  The point where the sum of the shot noise measurement imprecision (dotted line) and RSPN induced displacement fluctuations (black line) is minimized is termed the standard quantum limit (SQL)~\cite{Teufel09,Anetsberger10}.  Here, the displacement spectral density from RPSN at the mechanical resonance frequency, $\omega_m$, is $S_z^{SQL}(\omega_m)=\hbar/m \omega_m \Gamma_m$, where $m$ and $\Gamma_m$ are the resonator's mass and damping rate. This fundamental scale is equivalent to one half of the resonator's quantum mechanical zero point motion, $Z_{\mathrm{zp}}$. We also define $P^{SQL}$, the power required for a shot noise limited measurement imprecision of $S_z^{SQL}(\omega_m)$. Even with other mechanical noise sources present (e.g.~thermal motion - brown line) quantum backaction may still play an important role if the optical power, $P$, is sufficiently larger than $P^{SQL}$. 

	Whereas shot noise is a ubiquitous measurement limitation, experimental signatures of RPSN on solid objects have remained elusive.  Mechanical effects of photon recoil are routinely studied in atomic physics [\cite{StamperKurn12} and references therein], and a RPSN observation analogous to ours has been made using a dilute gas of ultracold atoms~\cite{Murch08}. A promising route to studying RPSN in solid objects is experiments that achieve high optomechanical coupling to high-frequency, small (nanometer to centimeter scale) mechanical resonators. Using such resonators, groups have initiated searches for RPSN~\cite{Tittonen99,Verlot11}, observed classical analogs of RPSN~\cite{Verlot09}, and predicted experimental signatures of RPSN~\cite{Heidmann97,Borkje10,Yamamoto10}.  Backaction on a nanomechanical resonator has also been observed using other measurement devices such as single electron transistors~\cite{Naik06}.  Resonators have even been cooled with electromagnetic radiation to near their motional ground state, illustrating the capacity for dominant coherent optical forces~\cite{Teufel11,Chan11,Verhagen12}.  In these experiments quantum backaction has been limited thus far to the scale of $Z_{\mathrm{zp}}$, whereas in this report we demonstrate a strong backaction heating effect from RPSN.  Note also, in near ground state cooling experiments correlations between shot noise and RPSN driven mechanical motion are an important component of the observed optical spectra~\cite{Khalili12} and are responsible for, for example, the sideband asymmetry observed in~\cite{SafaviNaeini12}.

\begin{figure}[ht]
	\hspace*{-0.2in}
		\includegraphics[scale=.81]{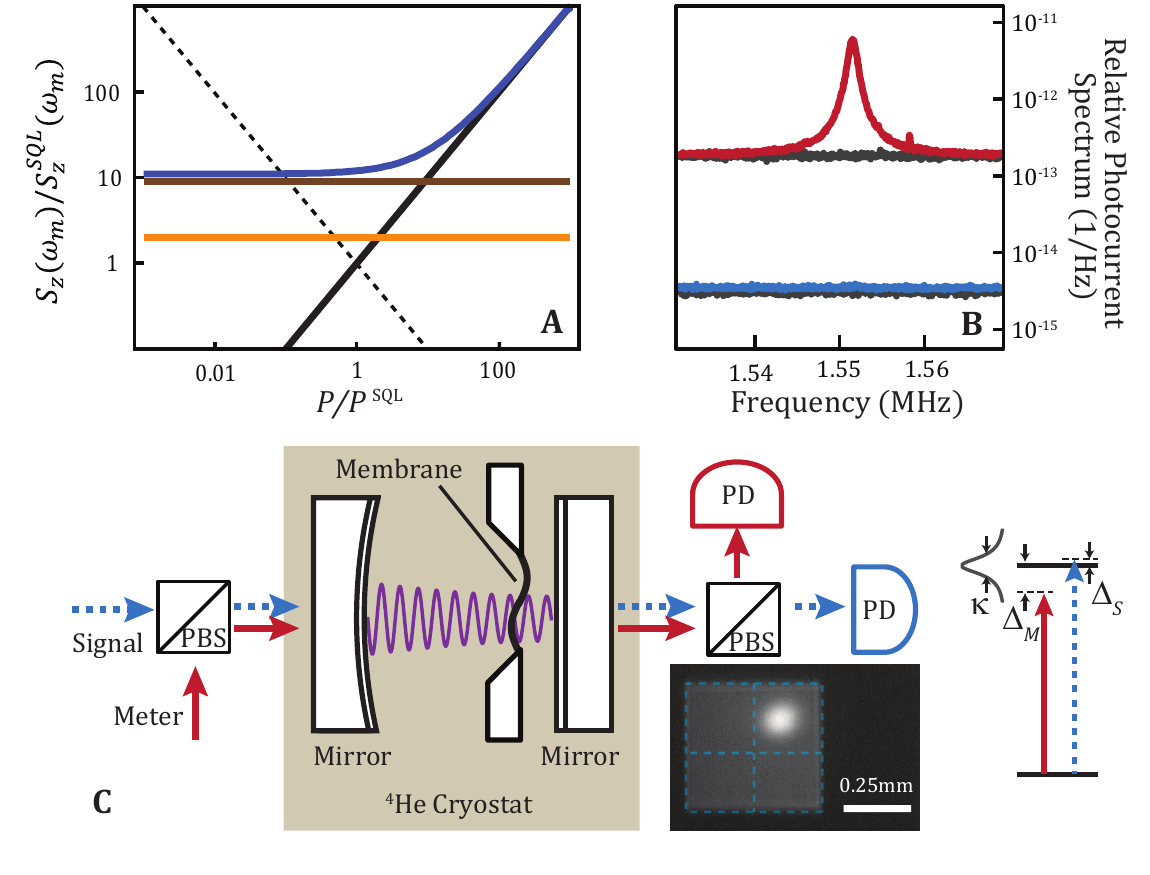}
	\caption{(A) Canonical picture of continuous position measurement.  RPSN (black), thermal motion (brown), and zero point motion (orange) combine to give the expected measurement result (blue). The dashed curve represents the effective displacement noise from the shot noise limited imprecision of an optical measurement. (B) Photocurrent spectra.  Plotted are the photocurrent spectral densities $S_{I_S}(\omega)/\bar{I}_S^2$ (blue), and $S_{I_M}(\omega)/\bar{I}_M^2$ (red), as well as the noise floors including detector noise and the dominant shot noise (gray). (C) Experimental setup.  Beams are combined and separated with polarizing beam splitters (PBS) and detected directly on photodetectors (PD).  The inset photograph shows an in-situ image of the square membrane and optical mode spot, with blue dashed lines indicating the nodes of the (2,2) mechanical mode. The inset diagram (right) shows laser-cavity detunings.}
	\label{fig:fig1}
\end{figure}

	Our optomechanical system consists of a silicon nitride membrane resonator inside of a Fabry-Perot optical cavity specially designed to operate at cryogenic temperatures [Fig.~1C and~\cite{Purdy12}].  Pioneering work by a group at Yale \cite{Thompson08} showed that membrane motion can be coupled to a cavity through a dispersive interaction, where the cavity resonance frequency shifts as the membrane moves along the optical standing wave.  This interaction imprints phase and amplitude modulation on transmitted laser light, allowing for readout of the membrane motion.  In conjunction, the laser applies an optical gradient force to the membrane, pushing it toward higher optical intensity.  Our membrane is a highly tensioned square plate with a 0.5 mm side length and 40 nm thickness, and an effective mass of about 7 ng.  We operate in a helium flow cryostat with the resonator at a base temperature of 4.9 K, where intrinsic mechanical linewidths, $\Gamma_0/2\pi$, are typically less than 1 Hz.  For the (2,2) mode oscillating at $\omega_m/2\pi=1.55$ MHz, we achieve a maximum single-photon optomechanical coupling rate $g/2\pi=16$ Hz.
	
	We use two laser beams derived from the same 1064 nm source, both coupled to the same spatial mode of the cavity, but with orthogonal polarizations [Fig.~1C and  \cite{Heidmann97,Verlot09}].  The half-planar, 5.1 mm long cavity has a full linewidth $\kappa/2\pi \sim 1$ MHz, which varies slightly with the membrane position. The high intensity ``signal'' beam is actively stabilized to the optical resonance.  This beam provides the RPSN, and its transmitted intensity fluctuations constitute a record, which is partially obscured by optical loss, of the optical force on the resonator.  The corresponding sensitive position measurement is wholly imprinted in the unrecorded phase quadrature.  Additional phase noise from fluctuations in the cavity-laser detuning precludes shot-noise-limited phase-quadrature detection \cite{Purdy12}.  The much weaker ``meter'' beam is tuned to the red of the optical resonance imprinting the resonator's displacement spectrum on its transmitted intensity. While its shot noise drive is much smaller, the meter beam provides optical Raman sideband cooling of the mechanical mode \cite{Marquardt07} to 1.7~mK.  The optical damping greatly eases the requirements on the signal-beam--cavity detuning due to both parametric instabilities at positive detuning and the contamination of cross correlation by thermal motion~\cite{Supp}\cite{Borkje10,Verlot11}, but does not change the sensitivity of the resonator to RPSN relative to thermal forces.     

	The effect of the optomechanical coupling on the resonator from a single laser \cite{Marquardt07,WilsonRae07}, or multiple beams \cite{Borkje10} has been well studied.  The resonator's mechanical susceptibility is modified to include optomechanical damping and frequency shifts from each laser.  Additionally, the effective phonon occupation, $n_m$, is modified.  The optomechanical damping cools the resonator;  RPSN increases the amplitude of motion.  In equilibrium a simple rate equation gives $n_{m}=(n_{\mathrm{th}} \Gamma_0+n_{S} \Gamma_{S}+n_{M}\Gamma_{M})/\Gamma_{m}$.  Here $n_{\mathrm{th}}$ is the thermal phonon occupation; $n_S$ and $\Gamma_S$ ($n_M$ and $\Gamma_M$) are the effective bath temperature and optomechanical damping rate of the signal (meter) laser.  The total mechanical damping rate is $\Gamma_m=\Gamma_0+\Gamma_S+\Gamma_M$.   In our experiments $\Gamma_M\gg \Gamma_0,\, \Gamma_S$, while $\Delta_S\sim0$ and $N_S \gg N_M$, where $\Delta_S$, $N_S$ ($\Delta_M$, $N_M$) are the laser-cavity detuning and intracavity photon occupation of the signal (meter) beam.  RPSN dominates over thermal noise when the ratio $R_S=(C_S/n_{\mathrm{th}}) (1+(2\omega_m/\kappa)^2)^{-1}>1$, where $C_S=4 N_{S}g^2/\kappa \Gamma_0$ is the multiphoton cooperativity.  We are able to reach this high cooperativity regime ($C_S\sim10^6$) due to the small mass, weak intrinsic damping, and cryogenic environment of our resonator.

\begin{figure}[ht]
	\centering
		\includegraphics[scale=.77]{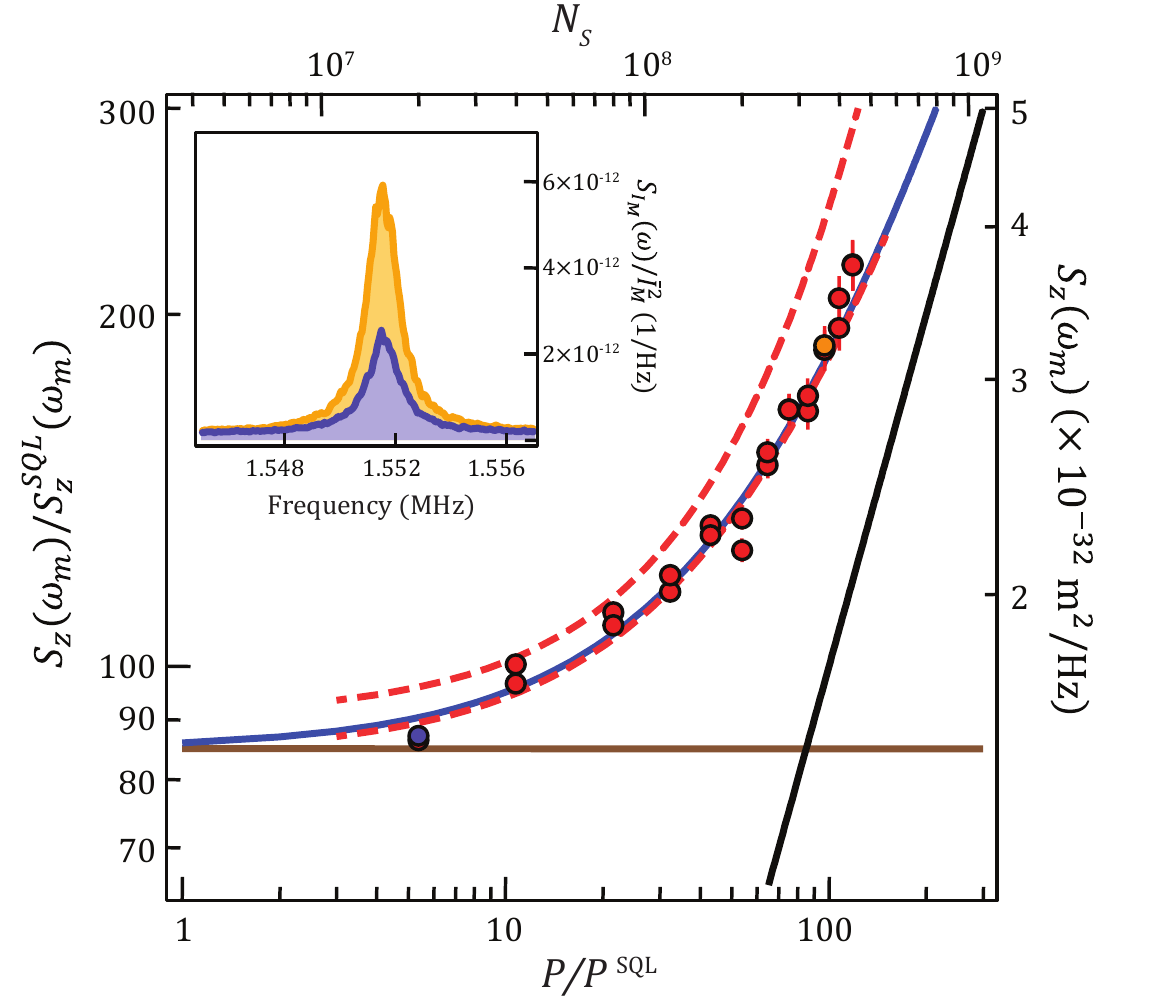}
	\caption{Displacement spectrum measurements.  Plotted are: measured peak displacement spectral density (circles), thermal contribution (brown), and expected RPSN contribution (black).  The blue curves represent the theoretical prediction for the sum of thermal motion and RPSN, and the dashed curves are bounds on theoretical estimates including systematic uncertainty in device parameters and the classical noise contribution.  Device parameters: $g/2\pi=16.1\pm0.3$ Hz, $\kappa/2 \pi=0.89$ MHz, $\Delta_S/2 \pi=2.0\pm 0.5$ kHz, $\Delta_M/2\pi=0.7$ MHz, $N_{M}=7.0\pm0.3\times 10^{6}$, $\omega_m/2 \pi=1.551$ MHz, $\Gamma_0/2 \pi =0.47$ Hz, $\Gamma_m/2 \pi =1.43$ kHz.  The inset shows transmission spectra for $R_{S}=0.056$ (blue) and $R_{S}=1.0$ (orange), with corresponding points in the main plot highlighted in blue and orange.}
	\label{fig:fig2}
\end{figure}

	The increase in phonon occupation resulting from RPSN is shown in Fig.~2.  The meter beam transmission spectrum,  $S_{I_M}(\omega)$ (Fig.~2 inset), shows a marked increase in spectral area or equivalently $n_m$ as the measurement strength is increased to where  $R_S\sim1$. Here the employed $N_S=3.6\times 10^8$ is equivalent to about 200 $\mu$W of detected optical power. The device shows good agreement with a theory of measurement backaction (Fig.~2 blue curves) that is based upon independently measured device parameters.  Because a separate meter beam is used to read out the mechanical motion, the measurement noise floor associated with these data is independent of the shot noise level of the signal beam as depicted by the dashed line in Fig.~1A.  The increased spectral density also includes a small contribution from classical radiation pressure noise.  Taking into account the thermal motion and classical laser intensity noise, we can attribute at least 40\% of the total displacement spectrum to RPSN at the maximum signal beam strength.  We have also measured similar backaction heating on another device with smaller $\Gamma_0$ and lower classical intensity noise (See Fig. S5).  The dashed curves of Fig.~2 represent bounds on the expected spectral densities accounting for systematic uncertainties in the device parameters and classical noise level~\cite{Supp}.  Another effect that might mimic RPSN is physical heating.  To test for physical heating, we monitor the temperature of a higher frequency, weakly optomechanically coupled mechanical mode where RSPN is negligible.  We do not observe a large response from this mode indicating the absorbed laser light causes a less than 10\% increase in the bath temperature (See Fig.~S4).

	We next examine the temporal correlations between the signal and meter beam photocurrents \cite{Heidmann97,Borkje10}.  We compute the spectrum of the two-time cross correlation function $S_{I_{SM}}(\omega)=\left<I_S^*(\omega)\, I_M(\omega)  \right> $, where $I(\omega)$ is the complex Fourier transform of the photocurrent $I(t)$, and the angle brackets represent an average over many realizations of the experiment.  Thermal and other ambient motion, as well as measurement noise uncorrelated to the radiation pressure drive are rejected by this technique, making it a powerful tool in understanding RPSN. In the limit $\Gamma_m \ll \kappa$, the correlation should reflect the Lorentzian response function of the optically damped resonator, driven by the locally white shot noise.  We show in Fig.~3A a cross correlation measurement and for reference, the product spectrum, $S_{I_S}(\omega)\times S_{I_M}(\omega)$.  $S_{I_S}(\omega)$ and $S_{I_M}(\omega)$ for these data are shown in Fig.~1B.  If the two beams are perfectly correlated the cross correlation and product spectra should coincide.  However, an uncorrelated measurement background, dominated by the meter's shot noise and thermal motion appear only on the product spectrum.  Additionally, the imperfect detection efficiency leads to a loss of correlation.  We measure a peak normalized correlation (the ratio of the red to black curve peaks in Fig.~3A) of $\mathcal{C}(\omega_m)=\left| S_{I_{SM}}(\omega_m) \right|^2/S_{I_S}(\omega_m) S_{I_M}(\omega_m)=0.14$.  An estimate, ignoring classical noise and assuming $\Delta_{S}=0$, is given by $\mathcal{C}(\omega_m)=R_S/(1+R_S) \times \kappa_R/\kappa \times \epsilon_{S}=0.15\pm0.02$, where $R_S/(1+R_S)=0.40\pm0.03$ is the fraction of $S_z(\omega_m)$ due to RPSN, $\kappa_R/\kappa=0.59$ is the fraction of the light through the output port, $\epsilon_{S}=0.63\pm0.03$ is the post-cavity detection efficiency.  By intentionally adding classical intensity noise much larger than shot noise to the signal laser, we demonstrate in Fig.~3B a (classical) normalized cross correlation that approaches unity.

\begin{figure}[ht]
	\hspace*{-0.15in}
		\includegraphics[scale=.79]{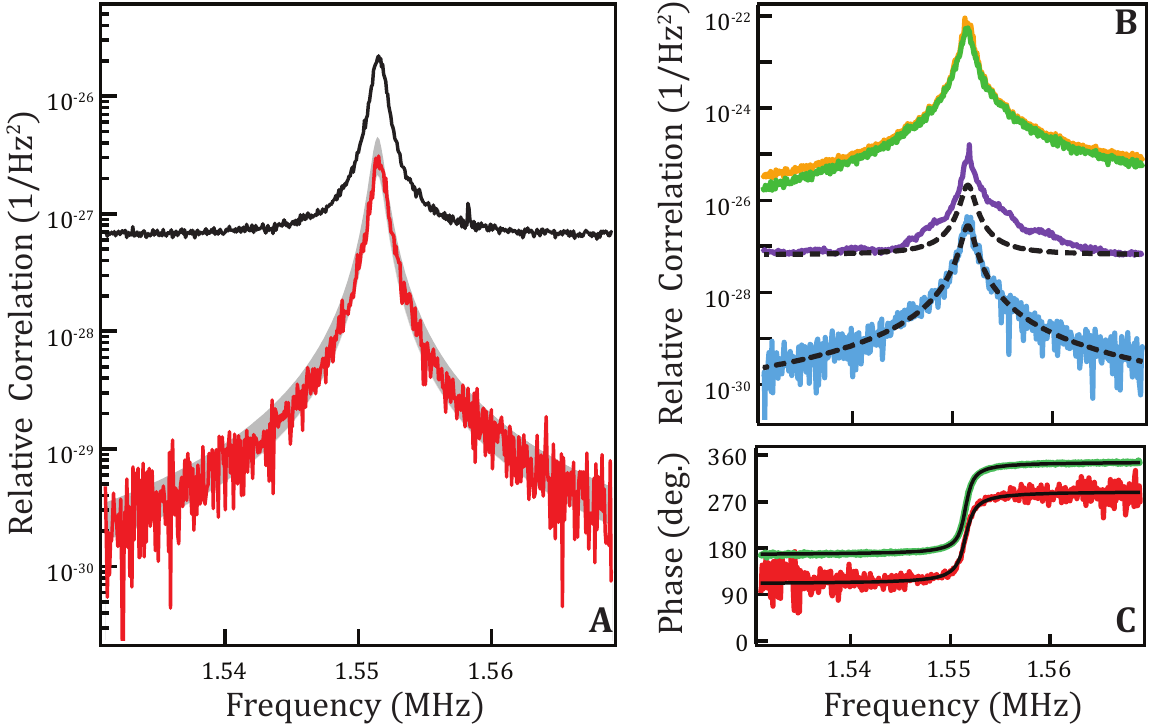}
	\caption{Cross correlation measurements.  (A) $|S_{I_{SM}}(\omega)/\bar{I}_S \bar{I}_M|^2$: measured (red), expected including systematic uncertainty (gray) and $S_{I_S}(\omega)/\bar{I}_S^2\times S_{I_M}(\omega)/\bar{I}_M^2$ (black).  Parameters are as listed in the Fig.~2 caption except $\Delta_S/2\pi=300\pm100$ Hz, $g/2\pi= 14.8\pm0.4$ Hz, and $N_{S}=3.2\times 10^8$.  The resolution bandwidth is 50 Hz.  (B) $|S_{I_{SM}}(\omega)/\bar{I}_S \bar{I}_M|^2$ (green) and $S_{I_S}(\omega)/\bar{I}_S^2\times S_{I_M}(\omega)/\bar{I}_M^2$ (orange) where classical intensity noise at the level of $\sim 40$ times shot noise, is added to the signal beam, raising the overall signal levels by the same factor.   $|S_{I_{SM}}(\omega)/\bar{I}_S \bar{I}_M|^2$  (blue) and $S_{I_S}(\omega)/\bar{I}_S^2\times S_{I_M}(\omega)/\bar{I}_M^2$ (purple) where membrane is driven with excess mechanical noise. Fits to the data of part (A) are displayed for reference (dashed black), showing that despite increased mechanical motion (purple curve above dashed curve), the correlation remains unchanged (blue curve coinciding with dashed curve). (C) Phase of the cross correlation with classical intensity noise on signal beam (green) and without (red).  Black curves are fits to the data.}
	\label{fig:fig3}
\end{figure}

	Figure 3C shows the phase of the correlation both with and without large classical intensity noise on the signal beam.  Both show the 180$^{\circ}$ phase shift expected from the mechanical response.  Importantly, we also expect a phase offset of $\arctan\left(2 \omega_m/\kappa\right)$ between the classical noise dominated drive and the shot noise dominated drive~\cite{Supp,Borkje10}.  Measurements of this phase offset imply that 75\% of the radiation pressure drive is from shot noise, in agreement with the directly measured classical noise range in $S_{I_S}(\omega)$.

If $\Delta_S$ is not zero, the cross correlation will be distorted.  Mechanical motion transduced directly onto $I_S$ may constructively or destructively add to the RPSN correlation depending on the sign of $\Delta_S$.  By fitting the correlation data to the expected lineshape~\cite{Supp}, we estimate $\Delta_S=0.0003\kappa$, implying only a 3\% contribution to $S_{I_{SM}}(\omega_m)$ from thermal motion.  We have also performed an experimental test to demonstrate the rejection of ambient motion from the cross correlation spectrum (Fig. 3B).  Here, we mechanically excite the membrane with a white-noise driven piezoelectric actuator (purple trace exceeds dashed curve), which also drives mechanical modes of the mirrors and supports, leading to extra modulation.  However, the cross correlation spectrum (blue trace) remains unchanged, equal to the unperturbed spectrum (dashed curve), implying very little of the ambient motion is transduced.

The cross correlation can also be viewed as evidence that we have made a quantum non-demolition (QND) measurement of the intracavity photon fluctuations of the signal beam \cite{Jacobs94,Heidmann97}. Here, the membrane acts as the measurement device, with its state of motion recording the photon fluctuations over the band of the mechanical resonance.  $\mathcal{C}$ is equivalent to a state preparation fidelity for a nonideal QND measurement \cite{Holland90}. Further, it has been shown that frequency-dependent ponderomotive squeezing of the signal beam quantum noise is possible \cite{Fabre94}, and has recently been demonstrated in an atomic gas cavity optomechanical system \cite{Brooks12}.  For our current laser configuration ($\Delta_S=0$), we do not expect to see squeezing in the detected amplitude quadrature.  However, our device parameters are sufficient to realize much stronger squeezing than has previously been demonstrated, limited mainly by optical loss.  Our observations open the door to realizing position measurement near the SQL if residual thermal noise and excess cavity-laser phase noise can be eliminated with improved devices or a colder base temperature.
\nocite{Wilson09}
\nocite{Sankey10}

\begin{acknowledgments}
We thank Pen-Li Yu for technical assistance and Konrad Lehnert's group for helpful discussions.  This work is supported by: the DARPA QuASAR program, the ONR YIP, and the JILA NSF PFC.  TP thanks the NRC for support.  CR thanks the Clare Boothe Luce foundation for support.
\end{acknowledgments}


\clearpage
\onecolumngrid
\begin{large}
Supplementary Materials for:
	\begin{center}
	\begin{bf}
	Observation of Radiation Pressure Shot Noise on a Macroscopic Object
	\end{bf}
	\end{center}
\end{large}

\begin{center}
T. P. Purdy, R. W. Peterson, and C. A. Regal
\end{center}
\begin{center}
	\begin{it}
	JILA, University of Colorado and National Institute of Standards and Technology,\\
	and Department of Physics, University of Colorado, Boulder, Colorado 80309, USA\\[3\baselineskip]
	\end{it}
\end{center}

\begin{bf}
\noindent Materials and Methods \\
\end{bf}
\\[1\baselineskip]
\noindent \underline{\smash{Theoretical Methods}}\\

Here we calculate the response of an optomechanical system to two independent laser driving fields.  Additionally, we compute the spectrum of the two time cross correlation of the photocurrents of the transmitted laser fields.  This formalism is used to model our membrane cavity optomechanical system, as well as our specific detection setup that incorporates direct photodetection of the transmitted intensity of both laser fields.  A schematic of the experiment is given in Fig.~S\ref{fig:Supp1}.
\begin{suppfigure}[ht]
	\centering
		\includegraphics[scale=.85]{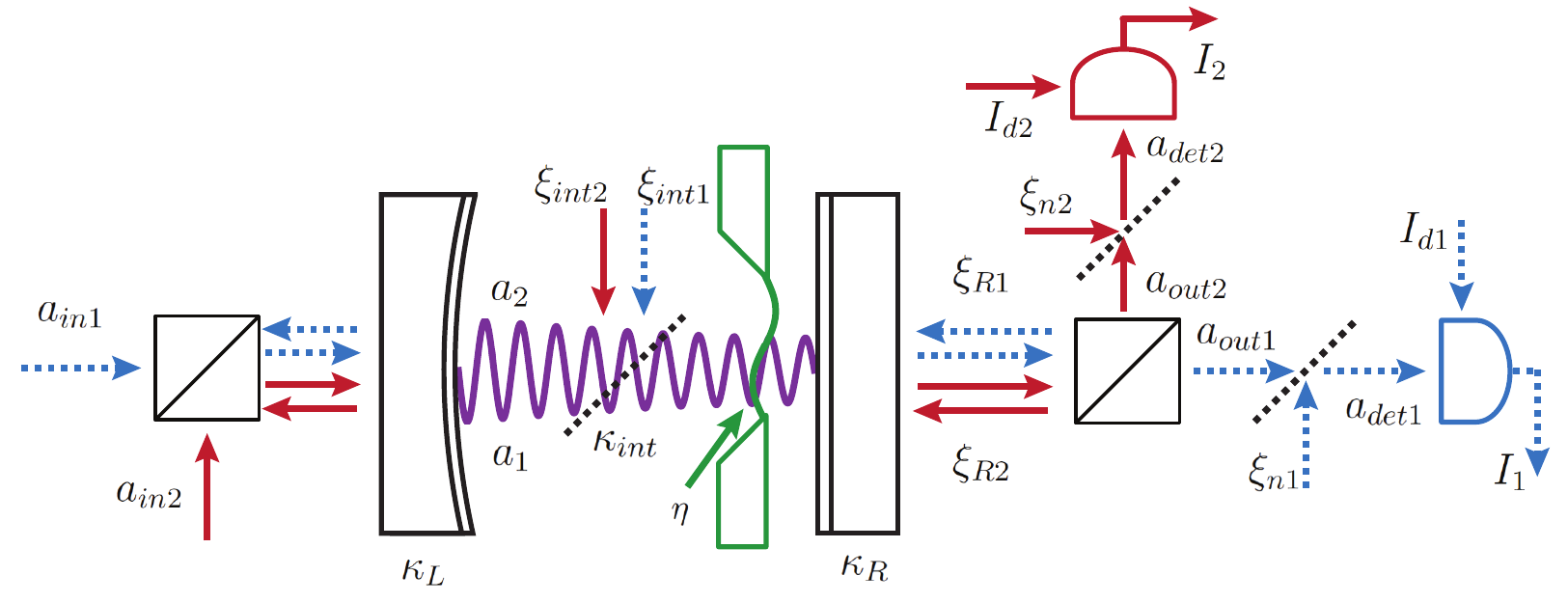}
	\caption{Cavity optomechanical probing and detection setup.  Two orthogonally polarized beams, laser 1 (dashed blue) and laser 2 (solid red) propagate through the system.  Boxes represent polarizing beam splitters.  Dashed lines indicate optical loss ports.  Various operators are labeled in accordance with the text.}
	\label{fig:Supp1}
\end{suppfigure}

\noindent \underline{\smash{Equations of Motion}}\\

We start from a Hamiltonian $H=H_0+H_{\kappa}+H_{\Gamma}$ describing the mechanical and optical evolution, dissipation and interactions ({\it 15}):
\begin{equation}
	H_0=\hbar \omega_m c^{\dag} c + \hbar \omega_c a^{\dag}_1 a_1 + \hbar \omega_c a^{\dag}_2 a_2+\hbar G_1 Z_{\mathrm{zp}} (c+c^{\dag}) a^{\dag}_1 a_1+\hbar G_2 Z_{\mathrm{zp}} (c+c^{\dag})a^{\dag}_2 a_2  \label{eq:Hamiltonian}
\end{equation}
where $\omega_m$ is the mechanical resonance frequency, $c$ is the annihilation operator for the mechanical resonator, $\omega_c$ is the cavity resonance frequency, and $a_1$, $a_2$ are annihilation operators for the two polarizations of the cavity mode.  $G_1$, $G_2$ are the optomechanical coupling constants, $Z_{\mathrm{zp}}=\sqrt{\hbar/2 m \omega_m}$ is the oscillator zero point motion, where $m$ is resonator's effective mass.  Single photon optomechanical coupling rates are $g_1=G_1 Z_{\mathrm{zp}}$, $g_2=G_2 Z_{\mathrm{zp}}$.  $H_{\kappa}$ represents the optical input and output coupling, and $H_{\Gamma}$ represents the coupling to the mechanical bath.  We will eventually specify the results of our calculation to identify the parameters of laser 1 with the signal beam ($_S$) and laser 2 with the meter beam  ($_M$) of the experiment.  The Heisenberg Langevin equations of motion are:
\begin{eqnarray}
	\dot{a}_1(t)&=&-\frac{\imath}{\hbar}\left[a_1(t),H_0 \right]-\frac{\kappa}{2} a_1(t)+\sqrt{\kappa_L} a_{\mathrm{inL1}}(t)+\sqrt{\kappa_{\mathrm{int}}} a_{\mathrm{int1}}(t)+\sqrt{\kappa_R} a_{\mathrm{inR1}}\nonumber\\
		\dot{a}_2(t)&=&-\frac{\imath}{\hbar}\left[a_2(t),H_0 \right]-\frac{\kappa}{2} a_2(t)+\sqrt{\kappa_L} a_{\mathrm{inL2}}(t)+\sqrt{\kappa_{\mathrm{int}}} a_{\mathrm{int2}}(t)+\sqrt{\kappa_R} a_{\mathrm{inR2}}\nonumber\\
		\dot{c}(t)&=&-\frac{\imath}{\hbar}\left[ c(t) ,H_0  \right]+\sqrt{\Gamma_0} \eta(t) \label{eq:EOM}
\end{eqnarray}
We have introduced a Langevin noise operator $\eta(t)$ to model the thermal mechanical bath, with mechanical decay rate $\Gamma_0$.  Additionally, we include input optical field operators for the three coupling channels of each cavity mode, input mirror $\kappa_L$, internal loss $\kappa_{\mathrm{int}}$, and output mirror $\kappa_R$. We assume the cavity couplings are identical for both optical modes.  By introducing the definitions for the optical field operators we find:

\[
	a_1(t)=(\bar{a}_1+d_1(t))e^{-\imath \omega_1 t},\quad a_2(t)=(\bar{a}_2+d_2(t))e^{-\imath \omega_2 t}
\]
\[
	a_{\mathrm{inL1}}(t)=(\bar{a}_{in1}+\xi_{L1}(t)+dx_1(t))e^{\imath \omega_1 t},\quad a_{\mathrm{inL2}}(t)=(\bar{a}_{in2}+\xi_{L2}(t))e^{\imath \omega_2 t}
\]
\[
	a_{\mathrm{inR1}}(t)=\xi_{R1}(t)e^{\imath \omega_1 t},\quad a_{\mathrm{inR2}}(t)=\xi_{R2}(t)e^{\imath \omega_2 t}
\]
\[
	a_{\mathrm{int1}}(t)=\xi_{\mathrm{int1}}(t)e^{\imath \omega_1 t},\quad a_{\mathrm{int2}}(t)=\xi_{\mathrm{int2}}(t)e^{\imath \omega_2 t}
\]
We allow for coherent field inputs from the left with real valued amplitudes $\bar{a}_{in1}=\left<a_{in1}\right>$, $\bar{a}_{in2}=\left<a_{in2}\right>$ and frequencies $\omega_1$, $\omega_2$, as well as vacuum noise inputs on all ports from the Langevin noise operators $\xi$, labeled by subscripts indicating the port and laser mode.  Classical input intensity noise on laser 1 is modeled with the real valued classical noise operator $dx_1$, which is measured in units relative to the shot noise level.  $\bar{a}_1$, $\bar{a}_2$ represent the complex valued classical amplitude of the intracavity field modes, while the operators $d_1$, $d_2$ represent the small classical and quantum fluctuations of the intracavity field. $z= Z_{\mathrm{zp}}(c+c^{\dag})-\bar{z}$ is the displacement operator, defined so the mechanical coordinate is centered around the optomechanically shifted equilibrium location $\bar{z}$.  $\Delta_1=\omega_c-\omega_1+G_1\bar{z}$ and $\Delta_2=\omega_c-\omega_2+G_2\bar{z}$ represent the input laser detuning from the optomechanically shifted cavity resonance. The equations of motion are linearized by neglecting small terms of order $d^2$, $d \times z$.  To solve the equations of motion we Fourier transform the Eqns. S\ref{eq:EOM} according to the following conventions: $f(\omega)\equiv\int^{\infty}_{-\infty}e^{\imath \omega t} f(t) dt$, $f^{\dag}(\omega)\equiv\int^{\infty}_{-\infty}e^{\imath \omega t} f^{\dag}(t) dt$, $\left(f^{\dag}(\omega)\right)^{\dag}=f(-\omega)$.

\begin{eqnarray}
d_1(\omega)&=&\chi_{c1}(\omega)\left(-\imath  G_1 \bar{a}_1 z(\omega)+\zeta_1(\omega)\right)\nonumber\\
d_2(\omega)&=&\chi_{c2}(\omega) \left(-\imath G_2 \bar{a}_2 z(\omega)+\zeta_2(\omega)\right)\nonumber\\
	\frac{z(\omega)}{Z_{\mathrm{zp}}}&=&\frac{1}{\mathcal{N}(\omega)}\Bigg(- 2 \omega_m \Big( \bar{a}_1^{*} g_1 \chi_{c1}(\omega)\zeta_1(\omega) + \bar{a}_1 g_1 \chi_{c1}^{*}(-\omega) \zeta^{\dag}_1(\omega) +  \bar{a}_2^{*} g_2 \chi_{c2}(\omega)\zeta_2(\omega) \nonumber \\
	&\quad& \qquad \qquad + \bar{a}_2 g_2 \chi_{c2}^{*}(-\omega) \zeta^{\dag}_2(\omega)\Big) +\sqrt{\Gamma_0}\left(\frac{\eta(\omega)}{\chi_m^{*}(-\omega)}+\frac{\eta^{\dag}(\omega)}{\chi_m(\omega)} \right) \Bigg)\nonumber
\end{eqnarray}
We have introduced the mechanical susceptibility $\chi_{m}(\omega)=\left( \Gamma_0/2-\imath(\omega-\omega_m) \right)^{-1}$, and the cavity susceptibilities for the two modes $\chi_{c1}(\omega)=\left(\kappa/2-\imath(\omega+\Delta_1)  \right)^{-1}$ and $\chi_{c2}(\omega)=\left(\kappa/2-\imath(\omega+\Delta_2)  \right)^{-1}$.  The optical noise operators are lumped into $\zeta_1(\omega)=\sqrt{\kappa_L}(\xi_{L1}(\omega)+dx_1(\omega))+\sqrt{\kappa_R} \xi_{R1}(\omega) + \sqrt{\kappa_{\mathrm{int}}} \xi_{\mathrm{int1}}(\omega)$ and $\zeta_2(\omega)=\sqrt{\kappa_L}\xi_{L2}(\omega)+\sqrt{\kappa_R} \xi_{R2}(\omega) + \sqrt{\kappa_{\mathrm{int}}} \xi_{\mathrm{int2}}(\omega)$.  The function \\$\mathcal{N}(\omega)=(\chi_m(\omega)\chi_m^*(-\omega))^{-1}- 2 \imath\omega_m Z_{\mathrm{zp}}^2 \left(|\bar{a}_1|^2 G_1^2 \Pi_1(\omega)+ |\bar{a}_2|^2 G_2^2 \Pi_2(\omega)\right)$ modifies the mechanical susceptibility and $\Pi_1(\omega)=\chi_{c1}(\omega)-\chi_{c1}^*(-\omega)$, $\Pi_2(\omega)=\chi_{c2}(\omega)-\chi_{c2}^*(-\omega)$.

	The two-sided mechanical displacement spectrum $S^{(2)}_{z}(\omega)=\left<z(-\omega) z(\omega)\right>$ may then be calculated from the expectation values of the operator pairs:
\[
 \left<\xi_{L1}(-\omega) \xi_{L1}^{\dag}(\omega)\right>=\left<\xi_{\mathrm{int1}}(-\omega) \xi_{\mathrm{int1}}^{\dag}(\omega)\right>=\left<\xi_{R1}(-\omega) \xi_{R1}^{\dag}(\omega)\right>=1
\]
\[
 \left<\xi_{L2}(-\omega) \xi_{L2}^{\dag}(\omega)\right>=\left<\xi_{\mathrm{int2}}(-\omega) \xi_{\mathrm{int2}}^{\dag}(\omega)\right>=\left<\xi_{R2}(-\omega) \xi_{R2}^{\dag}(\omega)\right>=1
\]
\[
	\left<\eta(-\omega)\eta^{\dag}(\omega)\right>=n_{\mathrm{th}}+1,\quad \left<\eta^{\dag}(-\omega)\eta(\omega)\right>=n_{\mathrm{th}}
\]
\[
	\left<dx_1(-\omega) dx_1(\omega)\right>=B_1
\]
Here $n_{\mathrm{th}}$ is the thermal occupation of the mechanical oscillator, and all other expectation values of products of Langevin operators are zero. The classical intensity noise is assumed to be locally white and takes a value $B_1$ times that of shot noise.  Here we assume the Langevin operators as well as the displacement spectrum are delta function correlated i.e. $\left<\xi(-\omega')\xi(\omega)\right>=\delta(\omega-\omega')$ with assumed integration over $\omega'$ for the calculation of all experimentally relevant quantities.
\begin{eqnarray}
\frac{S_{z}^{(2)}(\omega)}{Z_{\mathrm{zp}}^2}=\frac{1}{|\mathcal{N}(\omega)|^2}\Bigg(&\Gamma_0&\left(\frac{n_{\mathrm{th}}+1}{|\chi_m(\omega)|^2}+\frac{n_{\mathrm{th}}}{|\chi_m(-\omega)|^2}\right)\nonumber\\
&+&4\omega_m^2\kappa |\bar{a}_1 g_1 \chi_{c1}(-\omega)|^2+4\omega_m^2 \kappa |\bar{a}_2 g_2 \chi_{c2}(-\omega)|^2\nonumber\\
 &+&4\omega_m^2 \kappa_L |\bar{a}_1 g_1 \left(\chi_{c1}(\omega)+\chi_{c1}^{*}(-\omega)\right)|^2 B_{1} \Bigg)\label{eq:Sz}
\end{eqnarray}
The displacement spectrum consists of four terms.  The first term represents the residual thermal motion of the optically cooled oscillator.  The second term is the displacement due to RPSN from laser 1.  Assuming laser 2 is responsible for the majority of the optical damping, the third term contains most of the oscillator's zero point motion, which can also be thought of as backaction from laser 2, as well as the small RPSN heating effect from laser 2.  (Note, in the limit of large laser 2 power, the effective oscillator temperature will saturate at the so called Doppler limit, $n_{\mathrm{min}}=(\kappa/4\omega_m)^2$, where optomechanical cooling is balanced by RPSN.)  The last term is the response to classical intensity noise on laser 1. 

	Given the classical and quantum fluctuations in the optical spectrum, we want to calculate the mechanical response. Let us assume $\Delta_1\sim0$, as is the case in the actual experiment.  Then $A^{\mathrm{sn}}=(\kappa_R \bar{a}^*\bar{a})^{-1}$ is the output relative intensity spectrum of laser 1 due to shot noise, and  $A^{\mathrm{cn}}=\kappa_L |\chi_{c1}(\omega)+\chi_{c1}^*(-\omega)|^2 B_1/(\bar{a}^*\bar{a})$ is the output relative intensity spectrum of laser 1 due to classical intensity noise.  The classical output noise reflects filtering by the Lorentzian cavity response, whereas full shot noise appears on the output light.  However, inside the cavity the shot noise intensity fluctuations are suppressed by the cavity Lorentzian ({\it 25}).  Thus we must treat the perceived level of classical and shot noise differently to correctly infer the mechanical response.  Using Eq. S\ref{eq:Sz} we find that $S_z^{\mathrm{sn}}/S_z^{\mathrm{cn}}=\kappa \kappa_R |\chi_{c1}(-\omega)|^2 A^{\mathrm{sn}}/ A^{\mathrm{cn}}$.  Here $S_z^{\mathrm{sn}}$ and $S_z^{\mathrm{cn}}$ are the contribution to the displacement spectrum from the shot and classical noise on laser 1, equal to the second and fourth terms of Eq. S\ref{eq:Sz} respectively.  Experimentally this means that even when the measured classical intensity noise at an output photodetector is only a few percent of the shot noise level, it may still represent a significant amount of radiation pressure drive compared to the shot noise drive.
\\[1\baselineskip]
\noindent \underline{\smash{Correlation Spectrum}}\\

We next turn to the computation of the photocurrent cross correlation spectrum, $S_{I_{12}}^{(2)}(\omega)=\\ \left<\left(I_1(-\omega)-\bar{I}_1\right)\left( I_2(\omega)-\bar{I}_2\right)\right>_s$, with the mean photocurrents $\bar{I}_1=\left<I_1(t)\right>$, $\bar{I}_2=\left<I_2(t)\right>$.  Since the photocurrents are classical commuting variables, it should be that $I_1(\omega) \times I_2(\omega)=I_2(\omega) \times I_1(\omega)$.  To ensure this classical property, we compute the symmetrized expectation value for $S_{I_{12}}$, defining $\left<f(\omega) g(\omega)\right>_s=(1/2)\left(\left<f(\omega) g(\omega)\right>+\left<g(\omega) f(\omega)\right>\right)$.   For comparison with experimental data we compute the one-sided power spectrum $S_{I_{12}}(\omega)=S_{I_{12}}^{(2)}(\omega)+S_{I_{12}}^{(2)}(-\omega)$.  The photocurrents are given by
\[
I_1(t)=\hbar \omega_1 \mathcal{R}_1 a^{\dag}_{\mathrm{det1}}(t)a_{\mathrm{det1}}(t)+I_{d1}(t), \quad I_2(t)=\hbar \omega_2 \mathcal{R}_2 a^{\dag}_{\mathrm{det2}}(t)a_{\mathrm{det2}}(t)+I_{d2}(t)
\]
where $I_{d1}$, $I_{d2}$ are photodetector dark currents and $\mathcal{R}_1=q_e / \hbar \omega_1$, $\mathcal{R}_2=q_e / \hbar \omega_2$ are the photodetection sensitivities (with $q_e$ the photoelectron charge).  $a_{\mathrm{det1}}$ and $a_{\mathrm{det2}}$ are the photon annihilation operators at the photodetector (see Fig.~S\ref{fig:Supp1}).  The fields at the detector are a combination of the transmitted cavity fields, $a_{\mathrm{out1}}$, $a_{\mathrm{out2}}$ and vacuum noise from the optical loss channels.  Using the definitions $a_{\mathrm{out1}}(t)=(\bar{a}_{\mathrm{out1}}+d_{\mathrm{out1}}(t))e^{\imath \omega_1 t}$,  $a_{\mathrm{out2}}(t)=(\bar{a}_{\mathrm{out2}}+d_{\mathrm{out2}}(t))e^{\imath \omega_2 t}$ to distinguish the small fluctuations from the large classical amplitude $\bar{a}_{\mathrm{out1}}=\left<a_{\mathrm{out1}}(t)\right>$, $\bar{a}_{\mathrm{out2}}=\left<a_{\mathrm{out2}}(t)\right>$, we find:
\[
a^{\dag}_{\mathrm{det1}}(t)a_{\mathrm{det1}}(t)=\epsilon_{1} \bar{a}^*_{\mathrm{out1}}\bar{a}_{\mathrm{out1}}+\epsilon_{1}(\bar{a}^*_{\mathrm{out1}}d_{\mathrm{out1}}(t)+\bar{a}_{\mathrm{out1}}d^{\dag}_{\mathrm{out1}}(t))+\sqrt{\epsilon_{1}(1-\epsilon_{1})}(\bar{a}^*_{\mathrm{out1}}\xi_{n1}(t)+\bar{a}_{\mathrm{out1}}\xi^{\dag}_{n1}(t))
\]
\[
a^{\dag}_{\mathrm{det2}}(t)a_{\mathrm{det2}}(t)=\epsilon_{2} \bar{a}^*_{\mathrm{out2}}\bar{a}_{\mathrm{out1}}+\epsilon_{2}(\bar{a}^*_{\mathrm{out2}}d_{\mathrm{out1}}(t)+\bar{a}_{\mathrm{out2}}d^{\dag}_{\mathrm{out1}}(t))+\sqrt{\epsilon_{2}(1-\epsilon_{2})}(\bar{a}^*_{\mathrm{out2}}\xi_{n2}(t)+\bar{a}_{\mathrm{out2}}\xi^{\dag}_{n2}(t))
\]
where $\xi_{n1}$, $\xi_{n2}$ are Langevin vacuum noise operators, and the detection efficiencies $\epsilon_{1}$, $\epsilon_{2}$ include the photodetector quantum efficiencies and propagation losses outside of the cavity.  Substituting in the above relations, we find:
\begin{equation}
\frac{S_{I_{12}}^{(2)}(\omega)}{\bar{I}_1 \bar{I}_2}= \left\langle\ \left( \frac{\bar{a}^*_{\mathrm{out1}} d_{\mathrm{out1}}(-\omega){}+\bar{a}_{\mathrm{out1}} d^{\dag}_{\mathrm{out1}}(-\omega)}{\bar{a}_{\mathrm{out1}}^{*} \bar{a}_{\mathrm{out1}}}\right)\left(\frac{\bar{a}^*_{\mathrm{out2}} d_{\mathrm{out2}}(\omega)+\bar{a}_{\mathrm{out2}} d^{\dag}_{\mathrm{out2}}(\omega)}{\bar{a}_{\mathrm{out2}}^{*} \bar{a}_{\mathrm{out2}}}\right)\right\rangle_s \nonumber
\end{equation}
where we have employed the fact that $I_{d1}$, $I_{d2}$, $\xi_{n1}$, $\xi_{n2}$ are all uncorrelated with each other.  Thus terms proportional to expectation values of products of these operators are zero, and $S_{I_{12}}^{(2)}(\omega)/\bar{I}_1\bar{I}_2$ becomes independent of $\epsilon_{1}$, $\epsilon_{2}$.  We note, however, the power spectra of the individual photocurrents will depend on $I_{d1}$, $I_{d2}$, $\epsilon_{1}$, $\epsilon_{2}$, so the ratio $S_{I_{12}}/S_{I_1} S_{I_2}$ does improve with increasing detection efficiency and a lower photodetector noise floor.  We can relate the intracavity photon operators to the output operators using the boundary conditions $\bar{a}_{\mathrm{out1}}=\sqrt{\kappa_R} \bar{a}_1$, $\bar{a}_{\mathrm{out2}}=\sqrt{\kappa_R} \bar{a}_2$, and $	d_{\mathrm{out1}}=\sqrt{\kappa_R} d_1-\xi_{R1}$, $d_{\mathrm{out2}}=\sqrt{\kappa_R} d_2-\xi_{R2}$.  Applying the solution to the equations of motion from above, we find:
\begin{eqnarray}\label{eq:s12sol}
	\frac{S_{I_{12}}^{(2)}(\omega)}{\bar{I}_1 \bar{I}_2}&= & \frac{-1}{\kappa_R |\bar{a}_{1}|^2 |\bar{a}_{2}|^2} \times \nonumber \\
	&\bigg( & \sqrt{\kappa_R} G_1 |\bar{a}_1|^2 \Pi_1(-\omega)  \sqrt{\kappa_R} G_2 |\bar{a}_2|^2 \Pi_2(\omega) \left<z(-\omega)z(\omega) \right>_s \nonumber\\
	&+ & \sqrt{\kappa_R} G_1 |\bar{a}_1|^2 \imath \Pi_1(-\omega) \bar{a}_2^{*} \sqrt{\kappa_L \kappa_R} \chi_{c2}(\omega)\left<z(-\omega) \xi_{L2}(\omega)  \right>_s \nonumber\\
	&+ & \sqrt{\kappa_R}G_1 |\bar{a}_1|^2 \imath \Pi_1(-\omega) \bar{a}_2^{*} \sqrt{\kappa_{\mathrm{int}} \kappa_R} \chi_{c2}(\omega)\left<z(-\omega) \xi_{\mathrm{int2}}(\omega)  \right>_s\nonumber \\	
	&+ & \sqrt{\kappa_R}G_1 |\bar{a}_1|^2 \imath \Pi_1(-\omega) \bar{a}^{*}_2(\kappa_R \chi_{c2}(\omega)-1) \left< z(-\omega) \xi_{R2}(\omega)\right>_s \nonumber \\
	&+ & \sqrt{\kappa_R}G_1 |\bar{a}_1|^2 \imath \Pi_1(-\omega) \bar{a}_2 \sqrt{\kappa_L \kappa_R} \chi_{c2}^{*}(-\omega)\left<z(-\omega) \xi_{L2}^{\dag}(\omega)  \right>_s \nonumber \\
	&+ & \sqrt{\kappa_R}G_1 |\bar{a}_1|^2 \imath \Pi_1(-\omega) \bar{a}_2 \sqrt{\kappa_{\mathrm{int}} \kappa_R} \chi_{c2}^{*}(-\omega)\left<z(-\omega) \xi_{\mathrm{int2}}^{\dag}(\omega)  \right>_s \nonumber \\
	&+ & \sqrt{\kappa_R}G_1 |\bar{a}_1|^2 \imath \Pi_1(-\omega) \bar{a}_2(\kappa_R \chi_{c2}^{*}(-\omega)-1) \left< z(-\omega) \xi_{R2}^{\dag}(\omega)\right>_s \nonumber\\
	&+&  \sqrt{\kappa_R}G_2 |\bar{a}_2|^2 \imath \Pi_2(\omega) \bar{a}_1^{*} \sqrt{\kappa_L \kappa_R} \chi_{c1}(-\omega) \left(\left<\xi_{L1}(-\omega) z(\omega)  \right>_s + \left< dx_1(-\omega) z(\omega) \right>_s \right) \nonumber \\
	&+& \sqrt{\kappa_R} G_2 |\bar{a}_2|^2 \imath \Pi_2(\omega) \bar{a}_1^{*} \sqrt{\kappa_{\mathrm{int}} \kappa_R} \chi_{c1}(-\omega)\left<\xi_{\mathrm{int1}}(-\omega) z(\omega)  \right>_s \nonumber\\
	&+& \sqrt{\kappa_R} G_2 |\bar{a}_2|^2 \imath \Pi_2(\omega) \bar{a}_1^{*} (\kappa_R \chi_{c1}(-\omega)-1) \left< \xi_{R1}(-\omega) z(\omega) \right>_s \nonumber \\
	&+& \sqrt{\kappa_R} G_2 |\bar{a}_2|^2 \imath \Pi_2(\omega) \bar{a}_1 \sqrt{\kappa_L \kappa_R} \chi_{c1}^{*}(\omega) \left(\left<\xi_{L1}^{\dag}(-\omega) z(\omega)  \right>_s + \left< dx_1(-\omega) z(\omega) \right>_s \right) \nonumber \\
	&+& \sqrt{\kappa_R} G_2 |\bar{a}_2|^2 \imath \Pi_2(\omega) \bar{a}_1 \sqrt{\kappa_{\mathrm{int}} \kappa_R} \chi_{c1}^{*}(\omega) \left<\xi_{\mathrm{int1}}^{\dag}(-\omega) z(\omega)  \right>_s \nonumber \\
	&+& \sqrt{\kappa_R} G_2 |\bar{a}_2|^2 \imath \Pi_2(\omega) \bar{a}_1 (\kappa_R \chi_{c1}^{*}(\omega)-1) \left< \xi_{R1}^{\dag}(-\omega) z(\omega) \right>_s \bigg) \label{eq:SI12}
\end{eqnarray}
The necessary operator expectation values are:

\[
\left<z(-\omega) \xi_{j2}(\omega)  \right>_s=\frac{-\omega_m \sqrt{\kappa_j}}{\mathcal{N}(-\omega)} g_2 Z_{\mathrm{zp}} \bar{a}_2 \chi_{c2}^{*}(\omega),\quad \mathrm{for}\,\, j\in\{L,\mathrm{int},R\}
\]

\[
\left<\xi_{j1}(-\omega) z(\omega)  \right>_s=\frac{-\omega_m \sqrt{\kappa_j}}{\mathcal{N}(\omega)} g_1 Z_{\mathrm{zp}} \bar{a}_1 \chi_{c1}^{*}(-\omega), \quad \mathrm{for}\, j\in\{L,\mathrm{int},R\}
\]

\[
\left< dx_1(-\omega) z(\omega) \right>_s=\frac{-2 \omega_m \sqrt{\kappa_L}}{\mathcal{N}(\omega)} g_1 Z_{\mathrm{zp}} \left(\bar{a}_1^* \chi_{c1}(\omega)+\bar{a}_1 \chi_{c1}^*(-\omega)  \right) B_1
\]
To understand which terms are relevant in the experiment for RPSN, we note that $\Pi_1(\omega)$ approaches zero as $\Delta_1$ approaches zero.  The first seven terms of Eq. S\ref{eq:SI12} then vanish under these conditions.  This leaves terms that are proportional to correlations between $z$, the resonator position, and vacuum noise operators $\xi_{L1}$, $\xi_{R1}$ which represent the shot noise optical driving force on laser 1.  In the limit where $\Delta_1=0$ and in the absence of classical intensity noise, the cross correlation takes a simple form $\frac{S_{I_{12}}^{(2)}(\omega)}{\bar{I}_1 \bar{I}_2}=2 \imath g_1 g_2 \omega_m \frac{\Pi_2(\omega) \chi_{c1}^{*}(\omega)}{\mathcal{N}(\omega)}$.  For classical intensity noise an additional term is: $2 \imath g_1 g_2 \omega_m \kappa_L \frac{\Pi_2(\omega) |\chi_{c1}(\omega)+\chi_{c1}^*(-\omega)|^2}{\mathcal{N}(\omega)} B_1$.  Note that in our case where $\Gamma_m \ll \kappa$, the classical and shot noise terms show similar functional form near $\omega=\omega_m$. Importantly, the shot noise driven term includes an extra phase shift of Arg$(\chi_{c1}^{*}(\omega_m))\sim\arctan(2 \omega_m/\kappa)$.  This additional phase shift allows one to experimentally distinguish the quantum versus classical origin of a radiation pressure drive.  As shown in Fig. S\ref{fig:Supp2}B, the phase offset in the cross correlation varies continuously as the classical noise level is increased relative to the shot noise level. 

	In the experimentally relevant limit where $\Delta_1\rightarrow \Delta_S$, which is much smaller than $\kappa$ but still nonzero, the first term of Eq. S\ref{eq:SI12} produces a correlation induced by the mechanical motion of the resonator imprinted onto both photocurrents.  Depending on the sign of $\Delta_1$ the correlation from thermal motion may add either constructively or destructively with the RPSN correlation, leading to a laser frequency dependent lineshape.  Example expected cross correlation lineshapes are presented in Fig.~S\ref{fig:Supp2}A.  Note that this effect is most pronounced near the mechanical resonance peak. Over a wide range of $\Delta_1$ the Lorentzian wings of the mechanical resonance are insensitive to the thermal motion and give an accurate representation of the RPSN correlation.  The range of $\Delta_1$ values over which this effect is relevant is governed by the total mechanical damping rate, $\Gamma_m$.  Thus the range of $\Delta_1$ where the correlation retains its Lorentzian profile can be increased into an experimentally accessible regime by increased damping from laser 2.
\begin{suppfigure}[ht]
	\begin{center}
		\includegraphics[scale=.8]{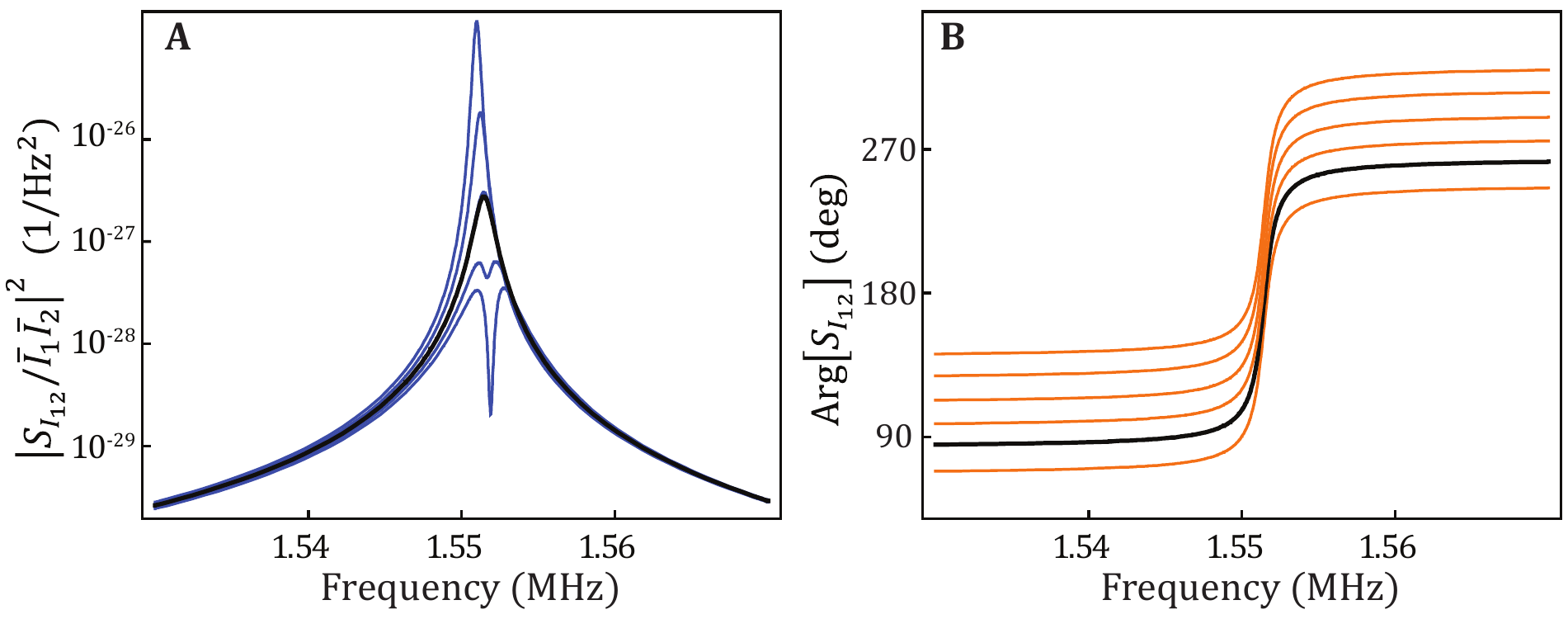}
	\end{center}
	\caption{Cross correlation spectra.  The magnitude squared (A) and the phase (B) of $S_{I_{12}}$ (described as $S_{I_{SM}}$ in the main text) are plotted for the parameters of the device described in the main text.  In (A) $\Delta_1$ is varied.  The tallest curve corresponds to $\Delta_1/2\pi=10$ kHz, and the lowest curve to -10 kHz. The black curve corresponds to 0.28 kHz, the value of the correlation measurement in Fig. 3A. In (B) the level of classical intensity noise relative to  shot noise on laser 1 is varied.  The bottom curve is pure quantum noise and the top curve is pure classical noise.  The black curve corresponds to a measured noise at the photodetector consisting of 91\% shot noise (75\% of radiation pressure drive from shot noise). This curve reflects the phase measurement in the correlation shown in Fig.~3C.}
	\label{fig:Supp2}
\end{suppfigure}
\\[1\baselineskip]
\noindent \underline{\smash{Optomechanical Instabilities}}\\

	One limit to the amount of optical power circulating in an optomechanical system is the onset of optomechanical bistability.  Here, the mean radiation pressure on the mechanical element causes a static displacement, which shifts the cavity resonance frequency by $\sim G_1 \bar{z}$.  When this shift becomes comparable to the cavity linewidth, there exist two stable values of the circulating power and $\bar{z}$ for a given input optical power, over some range of laser frequencies where the laser is nearly resonant with the cavity.  In this situation the system may be driven by small noise sources from one stable equilibrium to the the other in an uncontrolled fashion.  To some extent, such processes can be avoided by using active feedback, as is employed in the actual experiment, increasing the duration of stable operation.  To estimate the threshold for optomechanical bistability, we need to consider the net displacement of all the mechanical modes of our system with appreciable optomechanical coupling, not just the mode of interest for RPSN measurements.  The critical value of the intracavity photon occupation, $N^c_{1}$, above which bistable behavior is observed may be estimated for our membrane system by:
\[
N^c_{1}\sim\sum_{m,n}\frac{0.77\, \kappa\, m\, \omega_{(m,n)}^2}{\hbar\, G_{(m,n)}^2}
\]   
Here the sum is over mechanical eigenmode indicies $(m,n)$, and $\omega_{(m,n)}$, $G_{(m,n)}$ are the mechanical resonance frequency and optomechanical coupling constant for the mode $(m,n)$.  We estimate $G_{(m,n)}$ by measuring the location of the optical mode spot on the membrane, and then computing overlap integrals between the expected optical and mechanical mode functions for each mechanical mode.  For our measured parameters we obtain $N^c_{1}\sim3.5 \times 10^8$ which is comparable to the highest value of $N_{S}$ employed in the actual experiment.

	Another possible instability in the system arises when the net optical damping rate from all lasers is negative and larger in magnitude than the intrinsic damping rate, i.e. $\Gamma_m < 0$.  In this case noise on the mechanics is greatly amplified leading to a dynamical instability, where the system oscillates with increasingly large amplitude of motion, bounded only by the onset of nonlinearities.  Any mechanical mode may exhibit this behavior, depending on its coupling, frequency, and intrinsic damping rate.  To avoid this instability, we ensure that a net positive damping occurs for all of the modes by an appropriate choice of meter beam detuning.   \\[4\baselineskip]
\noindent \underline{\smash{Experimental Methods}}\\

	Here we describe in detail the calibration and operation of our optomechanical system.  More details about the construction of the system can be found in Ref. ({\it 23}). 
\\[1\baselineskip]
\noindent \underline{\smash{Experimental setup}}\\

	 The optical cavity consist of two mirrors, one flat, one with a 5 cm radius of curvature, both with approximately $1\times 10^{-4}$ fractional intensity transmission at the operating wavelength of 1064 nm and only a few $\times 10^{-6}$ scattering and absorption losses.  The mirrors are held at a separation L = 5.1 mm by an invar spacer.  With this geometry, the cavity is expected to support a Gaussian profile standing wave mode with a 1/$e^2$ intensity radius of 72 $\mu$m.  The flat mirror is attached to the invar spacer through a multilayer pizeoactuator that is used to precisely tune the overall optical path length of the cavity.  The cavity finesse without the membrane present is measured to be 31,000 via optical ringdown spectroscopy.

	A stoichiometric high-tensile-stress Si$_3$N$_4$ membrane ({\it32}) supplied by Norcada Inc. is placed inside the cavity as the mechanical element.  The membrane is square in shape, 0.5 mm on a side and is suspended on a silicon frame with dimensions 5 mm $\times$ 5 mm $\times$ 0.5 mm.  The membrane film is typically measured via ellipsometry to have a refractive index of 2.0 and film thickness of 40 nm, yielding a reflectivity of 10\% for 1064 nm light.  The mechanical modes that most strongly couple to the light are transverse ``drumhead'' modes, which in the high-tension limit, show sinousoidal displacement profiles, with motion out of the plane of membrane.  We label the modes with two indices $(i,j)$ denoting the number of antinodes of oscillation along each of the transverse direction.  For most of the experiments performed we focus on the (2,2) mode that has four antinodes, one in each corner of the membrane. This mode oscillates at a frequency $\omega_m/2\pi\sim 1.55$ MHz, which is twice the frequency of the fundamental (1,1) mode.  The intrinsic mechanical linewidth, which varies with temperature and mounting technique, is  measured in-situ, at a wavelength where the cavity finesse is low to avoid any optomechanical effects,  via mechanical ringdown spectroscopy, giving $\Gamma_0/2\pi =0.47$ Hz for the device shown in the main text.
	
	The membrane is positioned in the cavity, so that the optical mode spot is approximately aligned with one of the antinodes of the (2,2) membrane mode to attain the largest optomechanical coupling.  The membrane is located about 0.9 mm from the flat mirror, and its position can be finely tuned along the optical standing wave with another multilayer piezoactuator.  The location of the membrane is passively stable at the few tens of nanometers level during cryogenic operation.  In cooling from room temperature to cryogenic temperatures the membrane retains sub-milliradian alignment with the cavity optical axis.  Any residual angular misalignment distorts and displaces the optical mode shape.  By imaging the optical mode, as shown in the inset image of Fig.~1C, we can assess the level of angular deviation ({\it23}).  Angular misalignment couples near-degenerate optical modes, potentially dramatically changing the optomechanical coupling ({\it33}).  We empirically verify that no optical modes couple strongly to our mode of interest by looking for changes in the cavity transmission level and linewidth as the membrane and end mirror positions are scanned via the piezoactuators over a range encompassing the expected thermal drift. 
	
	 The effective cavity input and output coupling and  loss as well as the optomechanical coupling all vary with the location of the membrane along the optical axis.  We use optical ringdown measurements of the cavity linewidth when the membrane is placed at the operating point, in conjunction with a simple matrix model ({\it32}), to extract values for the cavity parameters.  We determine $\kappa$ the total cavity linewidth, $\kappa_L$ and $\kappa_R$, the input and output coupling of the cavity, and $\kappa_{\mathrm{int}}$, the internal cavity loss coupling.  $\kappa_{\mathrm{int}}$ is due mainly to membrane absorption and scattering as well as clipping of the optical mode by the membrane frame.  This analysis yields $\kappa/2\pi$=0.89 MHz, $\kappa_L=0.32\kappa$, $\kappa_R=0.59\kappa$, and $\kappa_{\mathrm{int}}=0.09\kappa$.

	A detailed diagram of the laser setup employed in the experiment is shown in Figure S\ref{fig:Supp3}.  A diode-pumped, monolithic non-planar ring oscillator type Nd:YAG laser from Innolight GmbH drives the entire experiment.  The laser is first spectrally filtered by passing through a 40 kHz Fabry-Perot optical cavity.   Next, the light is doubled-passed through an acousto-optical modulator (AOM) used for fast control of the laser frequency.  The laser is then split into two components that serve as the signal and meter beams.  Each component is passed through another AOM, which provides relative frequency control and independent intensity stabilization of the two beams.  The signal beam passes through an in-fiber electro-optical modulator (EOM) that adds frequency sidebands at 18 MHz for Pound-Drever-Hall frequency stabilization of the laser-cavity detuning.  At low frequencies the overall optical path length of the cavity is servoed to maintain the cavity resonance with the signal beam.  At higher frequencies, up to 100 kHz, the laser frequency is servoed via the common AOM.  With this servo system we are able to reduce the cavity-laser frequency fluctuations to the few kilohertz level.  The servo output is aggressively filtered to ensure there is no response near $\omega_m$.  To combat drift in the servo lock point, residual amplitude modulation from polarization drift in the EOM is actively canceled.  To accomplish this we apply a DC voltage to the EOM crystal to null the amplitude modulation measured by a photodetector sampling the beam after the EOM.  With this system, we typically see frequency offset drifts of less than 1 kHz over the tens of minutes times scale relevant to data taking.
	
\begin{suppfigure}[ht]
	\centering
		\includegraphics[scale=.7]{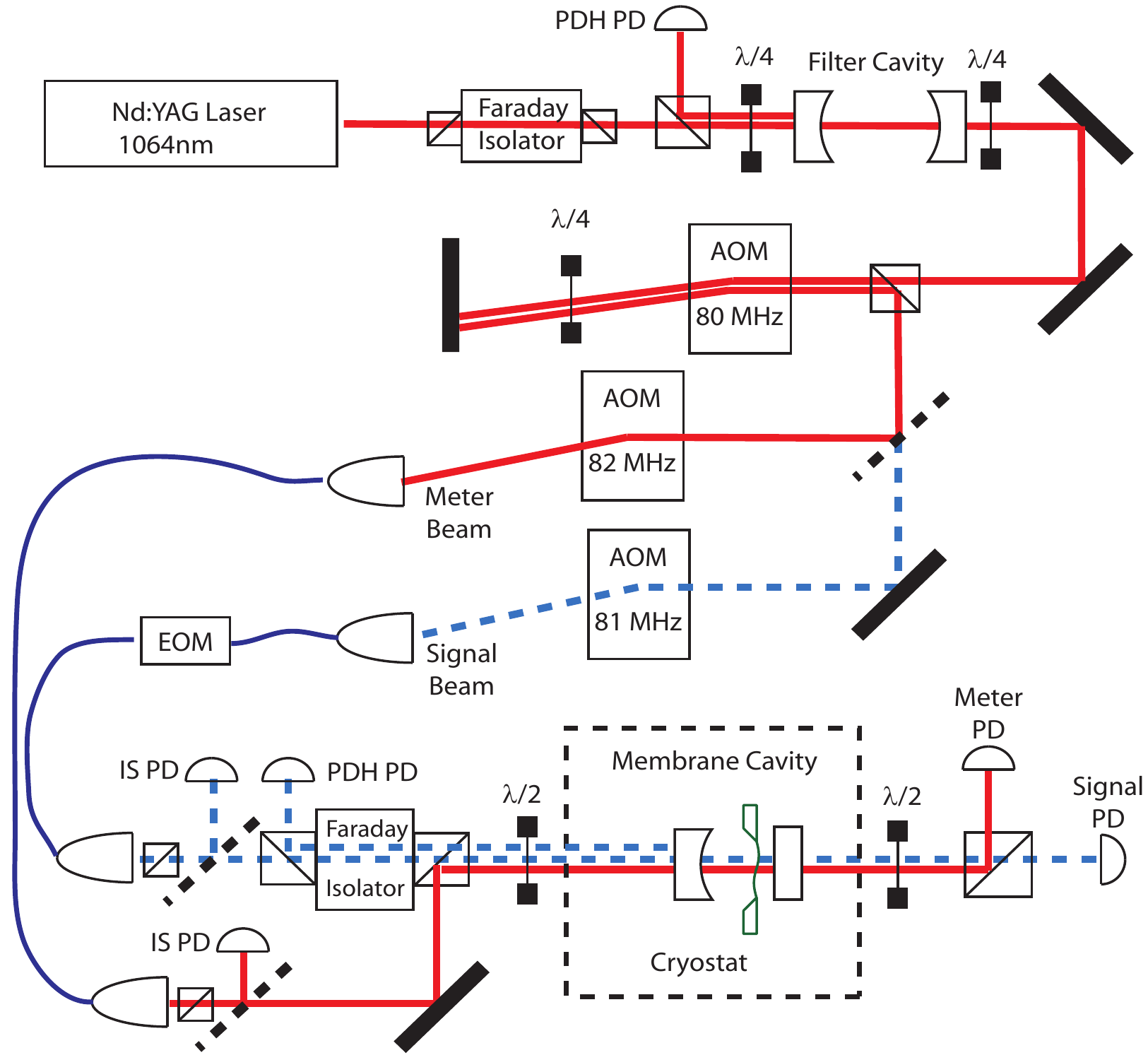}
	\caption{Detailed optical setup.  The signal beam (dashed blue) and meter beam (solid red) are derived from a single passively filter 1064 nm source.   Acousto-optical modulators (AOM) are used to shift the laser frequency.  An electro-optical modulator (EOM) is used to apply frequency sidebands for a Pound-Drever-Hall (PDH) frequency lock.  Light is detected at various points using photodetectors (PD).  Dashed black lines represent beam splitters, boxes represent polarizing beam splitters.  $\lambda$/4 and $\lambda/2$ are quarter-wave and half-wave retarders respectively.}
	\label{fig:Supp3}
\end{suppfigure}

	The signal and meter beams are combined on a polarizing beam splitter, and their polarizations are rotated to match the polarization eigenaxes of the slightly birefringent cavity ($\sim 400$ kHz birefringence splitting).  After the cavity, the two beams are split by another polarizing beam splitter and directed on to individual photodetectors.  We typically see less than $10^{-3}$ cross coupling between the two beams.  Because our polarization eigenmodes are unresolved by the cavity, it is possible that the two intracavity laser fields may beat against each other creating amplitude modulations of the intracavity field, if the lasers are not perfectly orthogonally polarized inside the cavity.  We ensure this amplitude modulation does not effect our measurements by setting the frequency difference of the two lasers to be at least several hundred kilohertz detuned from the mechanical resonance frequency.  Because both the mechanical and laser linewidths are much smaller than this frequency offset, the effect of the beating on the mechanics is minimized.  We estimate that the amplitude modulation at the highest optical powers employed should drive the resonator, in the worst case, by several tens of the amplitude of the zero point motion, but at a frequency far from the mechanical resonance.  The direct interference of the two lasers is evident in the photocurrent spectra, and is much larger than the optical modulation that results from the induced mechanical motion. It too is confined to a region of frequency far separated from the region used in the analysis.  We observe no evidence of interference near the mechanical resonance.  We also observe no residual background near the mechanical resonance in the cross correlation measurements that would be expected from the directly interfering frequency noise on the signal and meter beams, even in the absence of an optomechanical interaction.
	
	The optomechanical system is mounted onto the cold finger of a low-vibration $^4$He flow cryostat from Advanced Research Systems Inc., in which our device attains a base temperature of 4.9 K ({\it23}).  The cryostat has windows that allow direct free-space optical access to the cold sample region.  We include an additional copper radiation shield with small-aperture, fused-silica windows around our system, heat sunk to the cold finger to ensure our device thermalizes to the cryostat base temperature.  We are not inhibited by vibrations from cryogen flow in our system in part due to the low-vibration cryostat design made to eliminate boil off near the cold finger, and in part due to the near-monolithic design of our cavity ({\it23}) and high active feedback bandwidth.	
\\[1\baselineskip]
\noindent \underline{\smash{Calibration}}\\

	The optomechanical coupling of system is calibrated in two ways ({\it23}).  Because the signal and meter beams have the same spatial profile, we assume they share a common single-photon optomechanical coupling rate $g=g_S=g_M$ and coupling constant $G=g Z_{\mathrm{zp}}$.  At low signal laser power, the effect of the meter beam on the resonator can be predicted by calculations similar to those in the theory section, as is well documented in e.g.~({\it25}).   In the limit where $\Gamma_0,\,\Gamma_S \ll \Gamma_M$, the mechanical damping is given by $\Gamma_m=\Gamma_M=g^2 N_{M} \kappa \left(|\chi_{cM}(\omega_m)|^2-|\chi_{cM}(-\omega_m)|^2 \right)$.  Using this relation we can extract $g$ from $\Gamma_m |_{N_S=0}$ and a measurement of $N_{M}=\bar{I}_M/\left(q_e \epsilon_{M} \kappa_R \right)$  and the cavity parameters.   With this method we estimate $g/2\pi=16.4$ Hz.  Alternatively, we can use the thermal motion of membrane as a known displacement to calibrate $g$.   The effective temperature of the membrane mode is expected to be given by $T_{\mathrm{eff}}=T_{\mathrm{bath}} \Gamma_0/\Gamma_m$.  We measure $T_{\mathrm{bath}}$ from the silicon diode thermometer on the cryostat to be about 4.9 K.  Then $g^2=\left(S_{I_M}|_{N_S=0}(\omega_m)/\bar{I}_M^2\right) \, \Gamma_m^2 \hbar \omega_m/\left(8 |\Pi_2(\omega_m)|^2 k_b T_{\mathrm{bath}} \Gamma_0 \right)$, yielding $g/2\pi=15.8$ Hz.  For the cross correlation measurements, we measure a slightly smaller mechanical damping rate for the same laser parameters as compared the data in Fig.~1D.  Thus we estimate a slightly smaller value of $g/2\pi=14.8$ Hz for this data.  This difference can be accounted for by a small drift in the membrane position along the optical standing wave gradient between the two measurements.  
	
	
	 We estimate the detuning of the signal beam in a variety of ways.  For the data of Fig.~1D, we look for linear trends in both $\Gamma_m$ and the small change in the mechanical resonance frequency (the so called ``optical spring'' effect) as a function of $N_{S}$.   Using these trends, we estimate $\Delta_S/2\pi=2$ kHz.  For the cross correlation measurements we have taken more care to null $\Delta_S$.  Using the data of Fig.~3A, we fit the cross correlation to the functional form derived above, with $\Delta_S$ as a free parameter.  This fit yields $\Delta_S/2\pi=300\pm100$ Hz.  Below, in the analysis section, we show how to account for this coherent cooling or heating from the signal laser in order to accurately assess the contribution of RPSN from the mechanical displacement spectrum data.  Additionally, as discussed in the theory section, with nonzero $\Delta_S$, the cross correlation becomes distorted and contaminated by thermally driven mechanical motion.  
	
	The measured classical intensity noise on the signal beam after the cavity is always a small fraction of shot noise.  We assess its value and impact on the experiment in several ways.  Direct measurements of the total photocurrent power spectrum contain both shot noise and classical noise contributions, as well noise from photodetector dark current.  We calibrate the shot noise plus detector noise by illuminating the photodetector with a shot noise limited incandescent light source producing the same average photocurrent as the laser light used in the experiment.  We subtract the resulting photocurrent spectrum from $S_{I_S}(\omega)$ measured during the experiment, and expect the residual to be the classical noise.  The recorded shot noise spectral density is within a few percent of the predicted value, $2 q_e I_S$, based on a careful measurement of the photodetector transimpedence and digitization electronics frequency-dependent gain.  However, as the classical noise is typically less than 10\% of the shot noise level, this method is subject to significant error.  From these measurements we extract a classical noise power spectral density of less than -157 dBc/Hz.  We believe the majority of this noise arises from intensity noise imprinted on the laser from the electronics that drive the second AOM on the signal beam path as shown in Fig.~S\ref{fig:Supp3}.  Because the meter beam passes through an independent AOM with independent electronics, we do not expect any classical noise on the two lasers to be correlated. Further, thermally occupied mechanical modes of the mirrors and cavity support structure lead to noise in the cavity resonance frequency in the megahertz range.   However, because the signal beam is operated close to the cavity resonance, we do not expect a large effect from frequency to amplitude noise conversion by the cavity.  We experimentally confirm that we are insensitive to such effects, because our signal beam noise floor is, to good approximation, independent of $\Delta_S$ for $\Delta_S\ll\kappa$.

	  Classical laser intensity noise can also be detected and differentiated from shot noise by its effects on the membrane mechanics.  As pointed out in the theory section above, the phase of the cross correlation spectrum depends on classical versus quantum noise drive.  By fitting the phase of $S_{I_{SM}}(\omega)$, we estimate that 75\% of the correlation signal is due to RPSN, with most of the remainder accounted for by classical radiation pressure noise drive.  This value is within the range of the estimate made by direct photodetection of the classical noise.  Moreover, by taking into account the measured 25$^{\circ}$ phase shift in the photodetection electronics at frequencies near the mechanical resonance, we find excellent agreement between the absolute phase of the black theoretical curve of Fig.~S\ref{fig:Supp2}B and the red measured curve of Fig.~3C.
	  
	  Our main analysis of the RPSN data that is shown in Fig.~2 compares the data to an independent theory line.  Alternatively we can extract information about the classical noise level from fits to the RSPN data.  Here, we expect the increase in peak spectral density due to RPSN to scale linearly with the intracavity photon number.  However, for a constant classical relative intensity noise, the peak spectral density should scale quadratically with the intracavity photon number.  A three term polynomial fit to the data yields a constant term that is related to the thermal motion, a linear term related to RPSN, and a quadratic term related to classical radiation pressure.  This method indicates that $\sim75$\% of the increase in peak spectral density is due to RPSN for the highest $N_{S}$ value employed.       

\begin{suppfigure}[ht]
	\centering
		\includegraphics[scale=1]{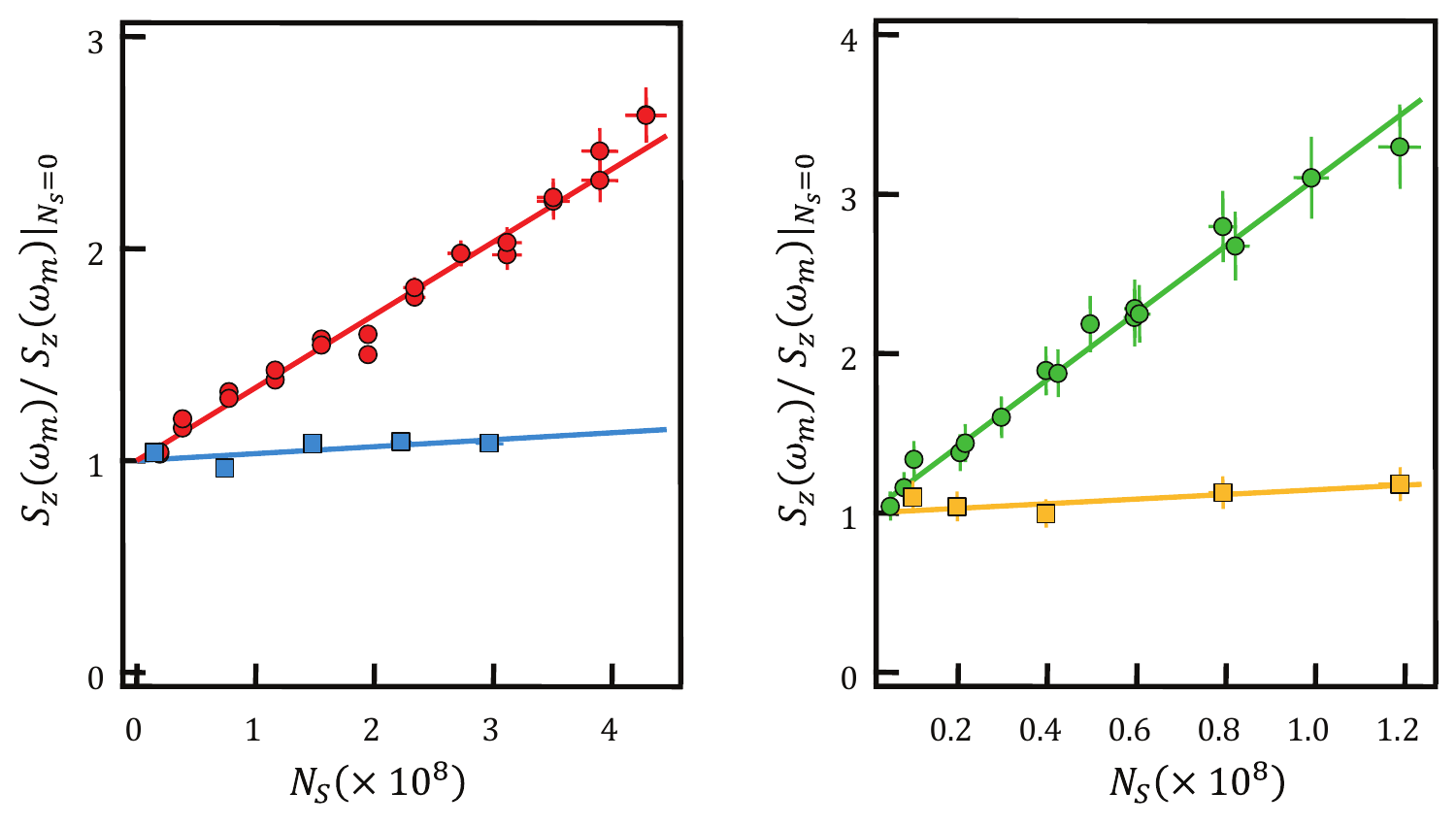}
	\caption{Response of multiple mechanical modes.  Left Panel: The peak displacement spectral density normalized to the peak displacement spectral density when $N_S=0$ is plotted for device from main text (2,2) mode (red circles) and (4,4) mode (blue squares).   Right Panel: Normalized peak spectral density for another membrane device (see Alternative device measurements subsection) is plotted for (2,2) mode (green circles) and (4,4) mode (yellow squares).  The (2,2) modes of both devices show a strong linear trend in response to RPSN, while the (4,4) modes show a relatively weak response, indicating that the bath temperature of the device is not affected by absorbed laser power.}
	\label{fig:Supp4}
\end{suppfigure}

	At large circulating optical power, it is possible that absorbed laser light will heat the membrane and increase $T_{\mathrm{bath}}$.  This effect, in the simplest case, would produce a linear increase in the peak displacement spectral density as function of optical power, the same scaling as RPSN.  To control for laser absorption heating, we look at the (4,4) membrane mechanical mode.  For the (4,4) mode the optomechanical coupling is weaker, the environmental coupling is stronger, and the oscillation is frequency higher.  Thus we expect only a few percent increase in the peak displacement spectral density due to RPSN at the highest drive power.  Any significant increase in the amplitude of motion of the (4,4) mode, we may then attribute to spurious heating effects.  The data of Fig. S\ref{fig:Supp4} show the response of the (2,2) and (4,4) mechanical modes for both the device used in the main text and a second device whose parameters are described in a later section.  There is only a weak linear trend in either of the (4,4) mode data sets.  This allows us to put an upper bound of 10\% for the increase in bath temperature at the largest $N_S$ employed for both devices, if we assume all of the mechanical modes are coupled to a common thermal bath.  
\\[1\baselineskip]
\noindent \underline{\smash{Data and Analysis}}\\

	For each data point in Fig.~2, we average the power spectra of a few hundred records of the photocurrents generated by signal and meter beams. The photocurrents are digitized with a two-channel, 16-bit digital oscilloscope card and a discrete Fourier transformation is applied to the digitized signals to generate power spectra.  Each record is 20 ms in length and is digitally sampled at $5\times 10^7$ samples per second for each laser power setting.  We compute $S_{I_M}(\omega)/\bar{I}_M^2$ and convert this into an effective displacement spectrum using the relation: $S_z(\omega)=\frac{S_{I_M}(\omega)}{\bar{I}_M^2}\frac{1}{|\Pi_M(\omega)|^2 G^2}$.  Over a region encompassing several mechanical linewidths around the mechanical resonance frequency, we fit each curve to a Lorentzian profile to extract the mechanical linewidth and peak spectral density.  Because $\Delta_S$ is a few kilohertz for this data, it provides an additional optomechanical damping and $\Gamma_m$ shows a linear trend for increasing $N_{S}$, with its value decreasing by about 20\% at the highest signal beam power.  By extrapolating to $N_{S}=0$, we find $\Gamma_m$ approaches $2 \pi \times 1.43$ kHz in the absence of the signal beam.  We then apply a small correction to the $S_z(\omega)$ data to remove the effect of the optical damping from signal beam.
	
	The dashed theory bounds of Fig.~2 are generated using the calibration methods discussed above.  The boundaries are generated by finding the extrema of the theoretical estimate of $S_z(\omega_m)$ using the two estimates for $g$ and the range of estimates for the classical intensity noise contribution.  The horizontal error bars on the data represent the systematic uncertainty in the conversion of mean photocurrent to $N_{S}$.  The vertical error bars represent statistical error in the measurement.          

	For the cross correlation data we take 1000 records of 20 ms in length digitally sampled at $2\times 10^7$ samples per second.  We simultaneously record both photocurrents.  We use vector averaging to compute the complex cross correlation spectrum, and from the same data also compute the scalar average of the power spectra of the two individual photocurrents.  We believe that 1000 averages is sufficient to converge the cross correlation spectrum over a region of several mechanical linewidths in spectral width.  As evidence, the blue data in Fig.~3B has converged to the same lineshape as the data in Fig.~3A, despite the added mechanical noise.  The boundary of the gray theory band of Fig.~3A is again generated as the extrema of the theoretical estimate using estimates for $g$ and the classical intensity noise level.    
\\[1\baselineskip]
\noindent \underline{\smash{Alternative device measurements}}\\

	We have also measured backaction heating on second membrane device with distinct parameters from the first device described in the main text.  Notably this second device has $\Gamma_0/2\pi=0.116$ Hz, about four times smaller than the first device.  Other device parameters are:  $g/2\pi=16.3\pm0.6$ Hz, $\kappa/2 \pi=1.17$ MHz, $\Delta_S/2 \pi=1.5\pm1.5$ kHz, $\Delta_M/2\pi=1.6$ MHz, $N_{M}=3.4\pm0.3\times 10^{6}$, $\omega_m/2 \pi=1.575$ MHz, $\Gamma_m/2 \pi =3$ kHz.  For the same $N_S$ the second device has $R_S$ about 5.0 times larger than the first device.  Thus RPSN effects show up on the second device at lower optical power.  The response of the two devices are compared in Fig.~S\ref{fig:Supp5}.  The second device also shows RPSN backaction heating confirming our measurements over a wider range of parameters, and with another of independently measured device parameters.  Because the second device is operated at overall lower optical power, the fractional classical laser intensity noise is also lower relative to shot noise for these measurements.  The inset of Fig.~S\ref{fig:Supp5} shows the measured transmitted signal laser noise floor at the photodetector always lies within a few percent of the measured shot noise floor for the second device. Referring this noise back to the intracavity level, we find at most 15\% of the displacement signal arises from classical laser noise. Taking into account the thermal motion and classical intensity noise, we can attribute at least 55\% of
the total displacement spectrum to RPSN at the maximum signal beam strength for the second device.

\begin{suppfigure}[ht]
	\centering
		\includegraphics[scale=1]{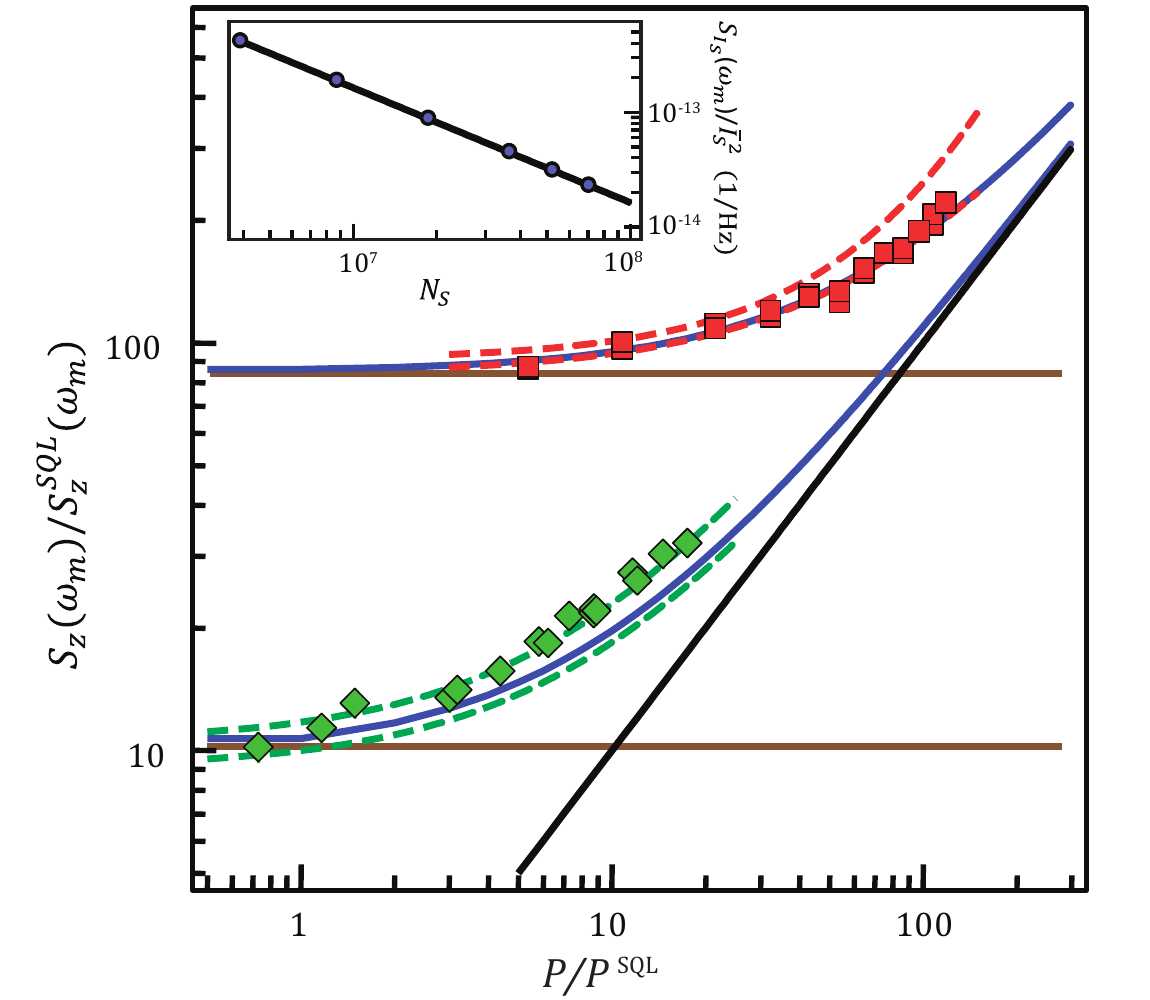}
	\caption{Comparison of backaction heating for two devices.  Plotted versus signal power $P$ are: Peak displacement spectral density for second device (green diamonds) and first device (red squares, reproduced from the main text for comparison), thermal contribution (orange), and expected RPSN contribution (black).   The blue curves represent the theoretical prediction for the sum of thermal motion and RPSN, and the dashed curves are bounds on theoretical estimates including systematic uncertainty in device parameters and classical noise contribution.  Inset shows measured spectral density for the signal beam photocurrent near $\omega_m$ (blue circles) for the second device and measured shot noise floor (black).}
	\label{fig:Supp5}
\end{suppfigure}

\end{document}